\documentclass[conference,a4paper]{IEEEtran}
\usepackage{xcolor}
\usepackage{balance}

\ifCLASSINFOpdf
  \usepackage[pdftex]{graphicx}
\else
  \usepackage[dvips]{graphicx}
\fi

\ifCLASSOPTIONcompsoc
 \usepackage[caption=false,font=normalsize,labelfont=sf,textfont=sf]{subfig}
\else
 \usepackage[caption=false,font=footnotesize]{subfig}
\fi

\usepackage{amsmath}
\usepackage{amsfonts}
\usepackage{amssymb}
\usepackage{acronym}
\usepackage{mathtools}                      % allows mulit-column cases*
\usepackage{subcaption}
\usepackage{bm}                             % bold math symbols
\usepackage{mathrsfs}
\usepackage{dsfont}

\acrodef{SVD}{singular Value decomposition}
\acrodef{LoS}{line-of-sight}
\acrodef{MIMO}{multiple-input multiple-output}
\acrodef{HoloS}{holographic surface}
\acrodef{DoF}{degrees of freedom}
\acrodef{eDoF}{effective degrees of freedom}
\acrodef{6G}{sixth-generation}
\acrodef{EM}{electromagnetic}
\acrodef{PSWF}{prolate spheroidal wave function}

\makeatletter
\newcommand{\mathleft}{\@fleqntrue\@mathmargin0pt}
\newcommand{\mathcenter}{\@fleqnfalse}
\makeatother

\begin{document}
%
% paper title
% Titles are generally capitalized except for words such as a, an, and, as,
% at, but, by, for, in, nor, of, on, or, the, to and up, which are usually
% not capitalized unless they are the first or last word of the title.
% Linebreaks \\ can be used within to get better formatting as desired.
% Do not put math or special symbols in the title.
\title{Degrees of Freedom of Holographic MIMO -- Fundamental Theory and Analytical Methods}

% author names and affiliations
% use a multiple column layout for up to three different
% affiliations
\author{\IEEEauthorblockN{
Juan Carlos Ruiz-Sicilia\IEEEauthorrefmark{1},   % 1st author, 1st affiliations
Marco Di~Renzo\IEEEauthorrefmark{1}\IEEEauthorrefmark{2},   % 2nd author, 2nd affiliations
Placido Mursia\IEEEauthorrefmark{3},    % 3rd author, 3rd affiliations
Vincenzo Sciancalepore\IEEEauthorrefmark{3},      % 4th author, 4th affiliations
Merouane Debbah\IEEEauthorrefmark{4}
}                                     % ...
%\\
%\IEEEauthorblockA{\IEEEauthorrefmark{1}% 1st affiliations
% Laboratoire des Signaux et Systémes, CNRS, Université Paris-Saclay, Gif-sur-Yvette, France}
 \IEEEauthorblockA{\IEEEauthorrefmark{1}% 1st affiliations
 Université Paris-Saclay, CNRS, CentraleSupélec, Laboratoire des Signaux et Systèmes, 91192 Gif-sur-Yvette, France} 
 \IEEEauthorblockA{\IEEEauthorrefmark{2}King's College London, Department of Engineering -- Centre for Telecommun. Research, WC2R 2LS London, UK} 
\IEEEauthorblockA{\IEEEauthorrefmark{3}% 2nd affiliations
NEC Laboratories Europe GmbH, Heidelberg, Germany}
\IEEEauthorblockA{\IEEEauthorrefmark{4}% 3rd affiliations
KU 6G Research Center, Khalifa University of Science and Technology, P.O. Box 127788, Abu Dhabi, UAE}
 \IEEEauthorblockA{ \emph{Corresponding author: juan-carlos.ruiz-sicilia@centralesupelec.fr} \vspace{-0.25cm}} 
}

% conference papers do not typically use \thanks and this command
% is locked out in conference mode. If really needed, such as for
% the acknowledgment of grants, issue a \IEEEoverridecommandlockouts
% after \documentclass

% use for special paper notices
%\IEEEspecialpapernotice{(Invited Paper)}

% make the title area
\maketitle

% As a general rule, do not put math, special symbols or citations
% in the abstract
\begin{abstract}
Holographic multiple-input multiple-output (MIMO) is envisioned as one of the most promising technology enablers for future sixth-generation (6G) networks. The use of electrically large holographic surface (HoloS) antennas has the potential to significantly boost the spatial multiplexing gain by increasing the number of degrees of freedom (DoF), even in line-of-sight (LoS) channels. In this context, the research community has shown a growing interest in characterizing the fundamental limits of this technology. In this paper, we compare the two analytical methods commonly utilized in the literature for this purpose: the cut-set integral and the self-adjoint operator. We provide a detailed description of both methods and discuss their advantages and limitations.
\end{abstract}

\vskip0.5\baselineskip

\section{Introduction}
Holographic \ac{MIMO} is considered a potential technology enabler for the future \ac{6G} wireless network \cite{HoloS2024,ngat}. This technology consists of an electrically large metasurface that can be modeled as a continuous aperture. A \ac{HoloS} is capable of inducing any current distribution on its surface and sensing any incident \ac{EM} field. Furthermore, this technology is energy and spectrum efficient due to the possibility of performing advanced signal processing computations in the \ac{EM} domain, as well as achieving a desirable spatial multiplexing and diversity trade-off with a limited number of radio frequency (RF) chains.

Spatial multiplexing refers to the transmission of multiple (not interfering) streams of data on the same time-frequency resource. Each data stream is termed a communication mode \cite{Miller_2019}, and the number of supported communication modes is the number of \ac{DoF} of the \ac{MIMO} link. Assuming that a \ac{HoloS} can be modeled as a continuous aperture, it can support an infinite number of communication modes. However, the finite bandwidth of typical wireless channels limits the number of highly coupled communication modes, i.e., those that carry a substantial fraction of the total transmit power. The number of highly coupled modes is referred to as the number of \ac{eDoF}. As a result, the theoretical characterization of the performance of a \ac{HoloS}-aided communication system boils down to the calculation of the number of \ac{eDoF} and the basis functions at the transmitter and receiver to jointly encode and decode multiple data streams without interference among them \cite{Miller}.

In recent years, the research community has shown a growing interest in charactering the \ac{eDoF} of holographic MIMO. Two main approaches can be applied to this purpose: the cut-set integral \cite{BucciBandwidth, BucciDoF, Franceschetti2011} and the self-adjoint operator \cite{Slepian1, Slepian2, Landau1975}. In this work, we discuss the advantages and disadvantages of both methods. Furthermore, we present notable research works focused on applying these methods to the analysis of holographic MIMO. 

Specifically, we introduce the system and signal models in Section II. In Section III, we describe the cut-set integral approach. In Section IV, we present the method based on the self-adjoint operator. In Section V, we compare both methods. Finally, conclusions are drawn in Section VI. 

\textit{Notation}: Bold lower and upper case letters represent vectors and matrices. Calligraphic letters denote sets. Script letters denote Hilbert spaces. $(\cdot)^*$ denotes the conjugate transpose. $\mathbb{R}^n$ denotes the space of real vectors of dimension $n$. $\mathbb{L}(\mathcal{S})$ denotes the Hilbert space of square-integrable functions defined in $\mathcal{S}$. $m(\mathcal{S})$ denotes the Lebesgue measure of $\mathcal{S}$. $j = \sqrt { - 1}$ is the imaginary unit. $\left\langle {f\left( {\bf{x}} \right),g\left( {\bf{x}} \right)} \right\rangle  = \int {f\left( {\bf{x}} \right){g^*}\left( {\bf{x}} \right)d{\bf{x}}}$ is the inner product of the functions ${f\left( {\bf{x}} \right)}$ and ${g\left( {\bf{x}} \right)}$. $f({\bf{x}}) * g({\bf{x}})$ is the convolution of $f({\bf{x}})$ and $g({\bf{x}})$. $\left| {\cdot} \right|$ and $|| {\cdot} ||$ denote the ${\ell}^2$-norm of a vector and a function, respectively. $\det \mathbf{(A)}$ is the determinant of matrix $\mathbf{A}$. $\mathds{1}_{\mathcal{P}}(\mathbf{x})$ denotes the indicator function over the set ${\mathcal{P}}$, i.e., $\mathds{1}_{\mathcal{P}}(\mathbf{x}) = 1$ if $\mathbf{x} \in \mathcal{P}$ and zero otherwise.

\section{Analytical Framework for DoF Analysis}
Let us consider a transmitting \ac{HoloS} and a receiving \ac{HoloS} located in $\mathbf{r}_{Tx} \in \mathcal{S}_{Tx}$ and $\mathbf{r}_{Rx} \in \mathcal{S}_{Rx}$, respectively. The transmitting \ac{HoloS} is characterized by a monochromatic current density distribution ${J}(\mathbf{r}_{Tx})\in \mathscr{X} = \mathbb{L}^2(\mathcal{S}_{Tx})$ that generates an electric field ${E}(\mathbf{r}_{Rx}) \in \mathscr{Y} = \mathbb{L}^2 (\mathcal{S}_{Rx})$ at the receiving HoloS, as follows:
\begin{equation}
\label{eq:sysmodel}
    {E}(\mathbf{r}_{Rx}) = ({K} {J})(\mathbf{r}_{Rx}) = \int_{\mathcal{S}_{Tx}} k(\mathbf{r}_{Rx}, \mathbf{r}_{Tx}) {J}(\mathbf{r}_{Tx}) d \mathbf{r}_{Tx}
\end{equation}
where $k(\mathbf{r}_{Rx}, \mathbf{r}_{Tx})$ denotes the propagation kernel from the transmitting HoloS to the receiving HoloS. In general, the operator $K$ is space-variant. 

Let us consider an orthonormal set of basis functions $\{{\phi}_n(\mathbf{r}_{Tx})\}$ and $\{{\psi}_n(\mathbf{r}_{Rx})\}$ to represent any signal in $\mathscr{X}$ and $\mathscr{Y}$, respectively. Then, ${J}(\mathbf{r}_{Tx})$ and ${E}(\mathbf{r}_{Rx})$ can be expressed as follows:
\begin{equation}
    {J}(\mathbf{r}_{Tx}) = \sum_{n=1}^{\infty} a_n {\phi}_n(\mathbf{r}_{Tx}), \quad\:
    {E}(\mathbf{r}_{Rx}) = \sum_{n=1}^{\infty} b_n {\psi}_n(\mathbf{r}_{Rx}) 
\end{equation}

\begin{figure}[!t]
    \centering 
    \vspace{0.1cm}
    \includegraphics[width=1.0\columnwidth]{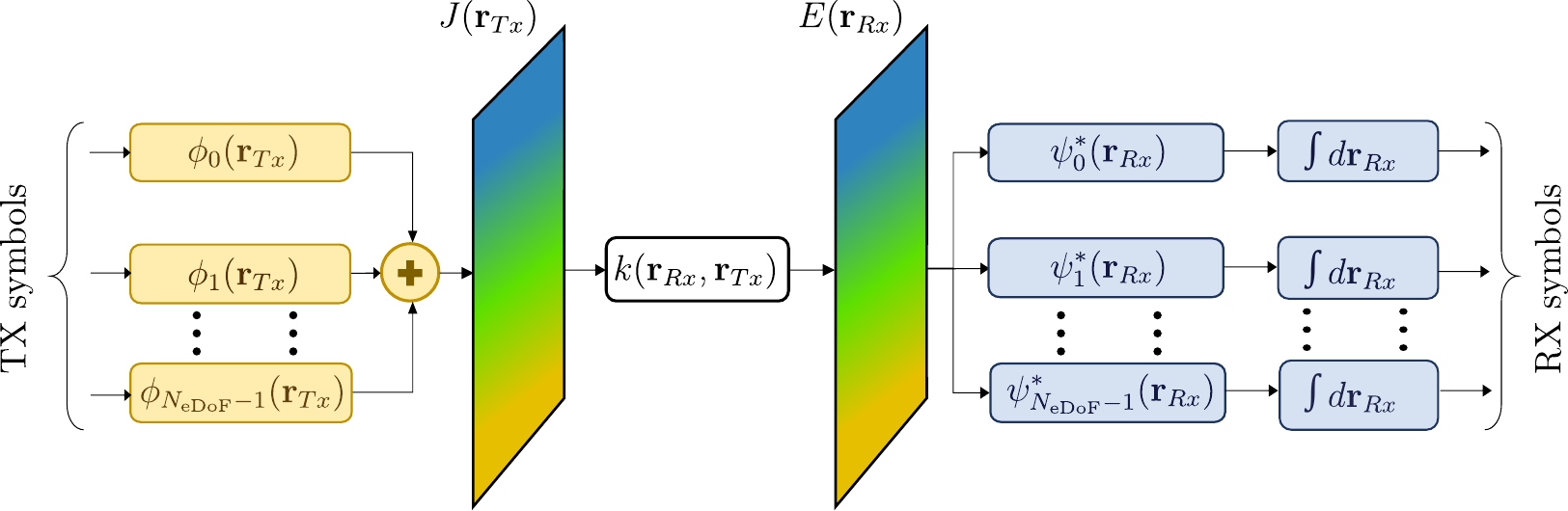}
    %\vspace{-0.4cm}
    \caption{Spatial multiplexing scheme}
    \label{fig:scheme}  \vspace{-0.7cm}
\end{figure}
As illustrated in Fig. \ref{fig:scheme}, a communication mode establishes a one-to-one correspondence between a basis function at the transmitter and a basis function at the receiver, so that the synthesis of the $n$-th basis function at the transmitter leads to the excitation of the $n$-th basis function at the receiver, while all the other basis functions are not excited. 

Let us assume that the kernel in \eqref{eq:sysmodel} is well-behaved, i.e.,
\begin{equation}
    \int_{\mathcal{S}_{Rx}} \int_{\mathcal{S}_{Tx}} ||{k}(\mathbf{r}_{Rx}, \mathbf{r}_{Tx})||^2 \, d\mathbf{r}_{Tx}d\mathbf{r}_{Rx} < \infty
\end{equation}
so that ${K}$ is bounded and compact, and is, therefore, a Hilbert-Schmidt operator. Then, the optimal basis functions can be obtained by solving the following equation \cite{Piestun:00}:
\begin{equation}
    s_n {\psi}_n (\mathbf{r}_{Rx}) = ({K} {\phi}_n)(\mathbf{r}_{Rx})
\end{equation}
where $s_n {\psi}_n(\mathbf{r}_{Rx})$ is the received electric field when exciting a current density ${J}(\mathbf{r}_{Tx}) = {\phi}_n(\mathbf{r}_{Tx})$ at the transmitting HoloS, and $s_n^2$ is the corresponding coupling intensity. 

Let us consider the adjoint operator of ${K}$, which is defined as ${K}^*:\mathscr{Y} \rightarrow \mathscr{X}$ whose kernel ${k}^*(\mathbf{r}_{Rx}, \mathbf{r}_{Tx})$ fulfills the condition $\left\langle({K} {J})(\mathbf{r}_{Rx}), {E})(\mathbf{r}_{Rx})\right\rangle = \left\langle{J}(\mathbf{r}_{Tx}), ({K}^* {E})(\mathbf{r}_{Tx})\right\rangle$. Then, we can also express the basis functions as the solution of the following two eigenproblems:
\begin{align}
    ({G}_{Tx} \phi_n)(\mathbf{r}_{Tx}) &= s_n^2 \phi_n(\mathbf{r}_{Tx}) \label{eq:eigen1}\\
    ({G}_{Rx} \psi_n)(\mathbf{r}_{Rx}) &= s_n^2 \psi_n(\mathbf{r}_{Rx}) \label{eq:eigen2}
\end{align}
where ${G}_{Tx} = {K}^* {K}$, defined in $\mathscr{X} \rightarrow \mathscr{X}$, and ${G}_{Rx} = {K} {K}^*$, defined in $\mathscr{Y} \rightarrow \mathscr{Y}$, are termed the self-adjoint operators. 

According to the spectral theorem, the eigenvalues $s^2_0 \geq s^2_1 \geq \dots \geq s^2_n \geq \dots \geq 0$ are non-negative and real, and they tend to $0$ as $n \rightarrow \infty$ \cite[Th. 4.15]{IntegralEq}. Also, the eigenvalues are upper-bounded by the operator norm of ${G}_{Rx}$, i.e., $||G_{Rx}||_{op} \geq s_n^2$ $\forall n$, which is defined as follows \cite[App. A]{IntegralEq}:
\begin{equation}
    ||G_{Rx}||_{op} = \sup_{||\psi(\mathbf{r}_{Rx})|| \leq 1} ||(G_{Rx} \psi)(\mathbf{r}_{Rx})||
\end{equation}

\subsection{Number of Effective DoF}
The number of eDoF is conventionally defined based on the effective dimensionality of the considered Hilbert space, in order to represent any element in $\mathscr{Y}$ up to a maximum approximation error that is defined according to the Kolmogorov $N$-width. This concept was first introduced in \cite{BucciDoF}. 

More precisely, let $\mathscr{Y}_N$ denote any $N$-dimensional subspace in $\mathscr{Y}$. The maximum approximation error that is obtained by using the subspace $\mathscr{Y}_N$ to approximate the image of the operator $K$ is given by
\mathleft
\begin{equation}
    D_{\mathscr{Y}_N} (K) = \sup_{||f(\mathbf{r}_{Tx})|| \leq 1} \inf_{g(\mathbf{r}_{Rx}) \in \mathscr{Y}_N} ||(K f)(\mathbf{r}_{Rx}) - g(\mathbf{r}_{Rx}) ||
\end{equation}
\mathcenter

The measure in (8) is known as the deviation. By considering the optimal $N$-dimensional subspace $\mathscr{Y}_N$ to minimize the deviation, we obtain the Kolmogorov $N$-width of the operator $K$ in the image space $\mathscr{Y}$, which is \cite[Ch. 2, Def. 1.1]{Kolmogorov}
\begin{equation}
    d_N(K) = \inf_{\mathscr{Y}_N \subseteq \mathscr{Y}} D_{\mathscr{Y}_N} (K)
\end{equation}

Based on the Kolmogorov $N$-width, we can define the number of \ac{eDoF} that correspond to an approximation error $\gamma$, as follows \cite{FranceschettiBook}:
\begin{equation}
\label{eq:Nedof_ini}
    N_{\mathrm{eDoF}, \gamma}(K) = \min \{N\, : \, d_N^2(K) \leq \gamma\}
\end{equation}

If ${K}$ is a Hilbert-Schmidt operator, it can be proved that $d_N(K) = s_N$ \cite{FranceschettiBook}. In simple terms, the deviation coincides with the square root of the $N$-th value of the coupling intensity. Then, \eqref{eq:Nedof_ini} becomes
\begin{equation}
\label{eq:Nedof}
    N_{\mathrm{eDoF}, \gamma}(K) = \min \{N\, : \, s_N^2 \leq \gamma \}
\end{equation}

From \eqref{eq:Nedof}, we can interpret the number of eDoF as the number of communication modes with a coupling intensity greater than or equal to the specified threshold $\gamma$.

\section{Cut-Set Integral}
In this section, we present the cut-set integral method for computing the number of \ac{eDoF}. This approach is also known as the spatial bandwidth method \cite{BucciBandwidth}. Also, we discuss the state of the art contributions to analyze holographic MIMO based on this approach.

\subsection{The Method in a Nutshell}
From sampling theory, it is known that the number of orthogonal functions i.e., the number of \ac{DoF}, to represent any signal $f(t)$ of finite duration $T$ and finite bandwidth $B$ is the Nyquist number $N_o = 2BT$ \cite{SlepianBandwidth}. It is important to remark that this is an approximation, since, according to the Heisenberg principle, it is not possible to have signals of finite duration and finite  bandwidth simultaneously. 

Similar to the relationship between time and frequency domains, we can consider the space and wavenumber domains, and we can apply similar arguments to the electric field ${E}(\mathbf{r}_{Rx})$. The wavenumber representation of $E(\mathbf{r}_{Rx})$ can be computed through the Fourier transform operator, as follows:
\begin{equation}
\label{eq:Fourier}
    (\mathcal{F} E)(\mathbf{k}_{Rx}) = \int_{-\infty}^\infty \int_{-\infty}^\infty E(\mathbf{r}_{Rx}) e^{-j(\mathbf{k}_{Rx}^* \mathbf{r}_{Rx})}\, d\mathbf{r}_{Rx}
\end{equation}

Similar to the frequency domain, the wavenumber domain provides an insightful representation of signals through the multidimensional Fourier transform. In this domain, signals are represented as a linear superposition of stationary plane waves with respect to the wavelength $\lambda$. A spectral component in the wavenumber domain corresponds to a plane wave with a specific direction of propagation \cite{wavenumberdomain}. Therefore, the wavenumber components are characterized by angles from which the transmitter can emit a planar wave to the receiver. 

When computing the electric field ${E}(\mathbf{r}_{Rx})$, the propagation kernel $k(\mathbf{r}_{Rx}, \mathbf{r}_{Tx})$ in (1) acts as a filter in the wavenumber domain, which limits the support of the transmitted signal in the wavenumber domain. Even considering a channel characterized by isotropic scattering, which provides the maximum wavenumber support, the bandwidth is limited to $W_{iso} = \pi k_0^2$ with $k_0 = 2\pi/\lambda$ \cite{Pizzo2022_chan}. In general, the support in the wavenumber domain depends on the points of observation on the receiving surface, i.e., $\mathcal{S}_{Rx}$. 

Let $W(\mathbf{r}_{Rx})$ denote the support in the wavenumber domain of a differential surface $d\mathbf{r}_{Rx}$. This is known as the local bandwidth. The contribution of $W(\mathbf{r}_{Rx})$ to the number of eDoF, or in simple terms to the Nyquist number considering the analogy with the time-frequency domain, is $dN_o = W(\mathbf{r}_{Rx})d\mathbf{r}_{Rx}/{(2\pi)^2}$, where the scaling term $1/(2\pi)^2$ is due to the definition of Fourier transform in  \eqref{eq:Fourier}. Integrating over the surface of the receiving \ac{HoloS}, we obtain the formulation in terms of cut-set integral (or spatial bandwidth) for the number of eDoF \cite{Franceschetti2011}, as follows:
\begin{equation}
\label{eq:cutset}
    N_o = \frac{1}{(2\pi)^2} \int_{\mathcal{S}_{Rx}} W(\mathbf{r}_{Rx}) \,d\mathbf{r}_{Rx}
\end{equation}

It is important to bear in mind that \eqref{eq:cutset} is, in general, an approximated result, since the supports in the spatial and wavenumber domains of any electric field ${E}(\mathbf{r}_{Rx})$ cannot be both finite simultaneously, similar to the time-frequency case \cite{SlepianBandwidth}. However, the authors of \cite{BucciDoF} proved that there is a very sharp transition between the magnitudes of $s^2_{N_o - 1}$ and $s^2_{N_o}$ when $N_o \rightarrow \infty$. More precisely, for an arbitrarily small $\gamma>0$, the following holds true:
\begin{equation}
    N_{\mathrm{eDoF}, \gamma}(K) = N_o \quad \text{when} \quad N_o \rightarrow \infty
\end{equation}

Based on \eqref{eq:cutset}, it is possible to compute the number of \ac{eDoF}, at least asymptotically, provided that the local bandwidth $W(\mathbf{r}_{Rx})$ is known. $W(\mathbf{r}_{Rx})$ depends on the channel operator ${K}$ under consideration, and analytical expressions are available for a few channel models, e.g., for isotropic channels \cite{Pizzo2022_chan} and \ac{LoS} channels \cite{BucciBandwidth, Dardari_2020, Ding2022, Ding2023}. 

%According to \cite{SlepianBandwidth}, the error in approximating ${E}(\mathbf{r}_{Rx})$ to bandlimited asymptotically tends to zero. This property can be exploited to compute the local bandwidth. For this purpose, let us define the filtered field ${E}_w(\mathbf{r}_{Rx})$ as follows
%
%\begin{equation}
 %   {E}_{\mathcal{Q}}(\mathbf{r}_{Rx}) = (\mathcal{F}^{-1} \mathds{1}_{\mathcal{Q}})(\mathbf{r}_{Rx}) *  {E}(\mathbf{r}_{Rx})
%\end{equation}

%Then, the wavenumber local bandwidth of ${E}(\mathbf{r}_{Rx})$ is $W(\mathbf{r}_{Rx}) = m(\mathcal{Q})$ where the set $\mathcal{Q}$ is optimized to fulfil  
%\begin{equation}
%    || {E}_{\mathcal{Q}}(\mathbf{r}_{Rx}) -  {E}(\mathbf{r}_{Rx})|| \rightarrow 0 \quad \text{when} \quad N_o \rightarrow \infty
%\end{equation}

%The local bandwidth $W(\mathbf{r}_{Rx})$ depends on the propagation 
\subsection{LoS Channels} 
Due to its relevance in the context of holographic communications, we focus on the LoS setting in this section. In this case, the kernel $k(\mathbf{r}_{Rx}, \mathbf{r}_{Tx})$ is given by the Green function
\begin{equation}
\label{eq:greenFunction}
     k(\mathbf{r}_{Rx}, \mathbf{r}_{Tx}) = \frac{j\eta \exp(-jk_0|\mathbf{r}_{Rx} - \mathbf{r}_{Tx}|)}{2\lambda |\mathbf{r}_{Rx} - \mathbf{r}_{Tx}|} 
\end{equation}
where $\eta$ is the intrinsic impedance in vacuum. 

%According to \cite{SlepianBandwidth}, the error in approximating ${E}(\mathbf{r}_{Rx})$ to be bandlimited asymptotically tends to zero. This property can be exploited to compute the local bandwidth. 

To compute the local bandwidth, the authors of \cite{BucciBandwidth} have introduced a filtered version of the received electric field, ${E}_{\mathcal{Q}}(\mathbf{r}_{Rx})$, which is defined as follows:
\begin{equation}
\label{eq:filteredField}
    {E}_{\mathcal{Q}}(\mathbf{r}_{Rx}) = (\mathcal{F}^{-1} \mathds{1}_{\mathcal{Q}})(\mathbf{r}_{Rx}) *  {E}(\mathbf{r}_{Rx})
\end{equation}
where $\mathcal{F}^{-1}$ denotes the inverse Fourier transform in (12).

The authors of \cite{BucciBandwidth} proved the asymptotic relation \eqref{eq:greenFunction}
\begin{equation}
    || {E}_{\mathcal{Q}}(\mathbf{r}_{Rx}) -  {E}(\mathbf{r}_{Rx})|| \rightarrow 0 \quad \text{when} \quad N_o \rightarrow \infty
\end{equation}
provided that all the wavemumber components $\mathbf{k}_{Rx}$ of ${E}(\mathbf{r}_{Rx})$ are contained in the set $\mathcal{Q}$.

Accordingly, the local bandwidth $W(\mathbf{r}_{Rx})$ evaluated at $\mathbf{r}_{Rx}$ coincides with the Lebesgue measure of the set spanned by varying $\mathbf{r}_{Tx}$ in $\mathcal{S}_{Tx}$, which is denoted by $m(\cdot)_{\mathbf{r}_{Tx} \in \mathcal{S}_{Tx}}$ and is defined as follows:
\begin{equation}
    W(\mathbf{r}_{Rx}) = m( \mathbf{k}_{Rx}(\mathbf{r}_{Rx}, \mathbf{r}_{Tx}))_{\mathbf{r}_{Tx} \in \mathcal{S}_{Tx}}
\end{equation}

The authors of \cite{BucciBandwidth} obtained a closed-form expression for the local bandwidth in (18) when the receiver is one-dimensional. The authors of \cite{Franceschetti2011} have discussed the generalization to two-dimensional receivers, which is the case of interest in the context of HoloS. The author of \cite{Dardari_2020} has recently introduced an approximated expression to compute the wavenumber components of ${E}(\mathbf{r}_{Rx})$ for two-dimensional receivers, which is given by
\begin{equation}
\label{eq:spectralComponent}
    \mathbf{k}_{Rx}(\mathbf{r}_{Rx}, \mathbf{r}_{Tx}) = k_0 [\hat{\mathbf{r}} - \hat{\mathbf{n}}(\hat{\mathbf{r}}^* \hat{\mathbf{n}})]
\end{equation}
where $\hat{\mathbf{r}} = (\mathbf{r}_{Rx} - \mathbf{r}_{Tx})/|\mathbf{r}_{Rx} - \mathbf{r}_{Tx}|$ and $\hat{\mathbf{n}}$ is the vector normal to the receiving \ac{HoloS}.

Based on (19), the authors of \cite{chen2024multiuser} have introduced an analytical framework to compute the local bandwidth that consists of two steps. First, the local bandwidth of a transmitting HoloS of (infinitesimal) size $d\mathbf{r}_{Tx}$ is computed as follows:
\begin{equation}
    d\mathbf{k}_{Rx}(\mathbf{r}_{Rx}, \mathbf{r}_{Tx}) = \text{det}(\mathbf{J}(\mathbf{r}_{Rx}, \mathbf{r}_{Tx})) d\mathbf{r}_{Tx}
\end{equation}
where $\text{det}(\mathbf{J}(\mathbf{r}_{Rx}, \mathbf{r}_{Tx}))$ is the Jacobian determinant of \eqref{eq:spectralComponent}, which is formulated analytically in \cite{chen2024multiuser}. Then, the local bandwidth is obtained by integrating (20) over the whole transmitting HoloS, as follows:
\begin{equation}
    W(\mathbf{r}_{Rx}) = \int_{\mathcal{S}_{Tx}} \text{det}(\mathbf{J}(\mathbf{r}_{Rx}, \mathbf{r}_{Tx})) d\mathbf{r}_{Tx}
\end{equation}

\subsection{Application to Holographic MIMO}
The cut-set integral has been widely used in the literature to compute the number of \ac{eDoF} of \acp{HoloS}  \cite{Dardari_2020, Ding2022, Ding2023, chen2024multiuser}. In \cite{Ding2022}, the authors characterized the number of eDoF between two linear holographic arrays. They obtained an expression for the number of \ac{eDoF} for any position, orientation, and size of the linear arrays. This work was further extended in \cite{Ding2023}, where the same authors introduced accurate closed-form approximations for the framework in \cite{Ding2022}. 

As far as \acp{HoloS} (i.e., two-dimensional holographic arrays) are concerned, \cite{Dardari_2020} introduced an approximated analytical expression to compute the number of \ac{eDoF} between two \acp{HoloS} of any size. Based on \cite{Dardari_2020}, the authors of \cite{chen2024multiuser} developed a general framework to compute the number of \ac{eDoF} in a multi-user holographic system, i.e., (21) was introduced.

\section{Self-Adjoint Operator}
In this section, we present the method based on the analysis of the self-adjoint operator in \eqref{eq:eigen2}, and summarize the available research works for application to holographic MIMO.

\subsection{Historical Perspective}
To account for the fundamental physical limit that a signal cannot have a finite support in the time and frequency domains (or equivalently in the spatial and wavenumber domains) simultaneously, the authors of \cite{Slepian1} computed the number of eDoF (for one-dimensional signals) by formulating a concentration problem. Conceptually, the approach is similar to the evaluation of the error in (17) based on a ``truncated'' version of the signal in (16). Subsequently, the authors of \cite{Slepian2} devised an analytical framework to compute the most concentrated orthogonal functions that minimize the approximation error induced by the formulation in terms of concentration. In \cite{Landau1975}, the author generalized this analysis for application to multidimensional signals, which includes HoloSs (i.e., two-dimensional spaces) as a special case. 

\subsection{The Method in a Nutshell}
The framework proposed in \cite{Landau1975} requires that the self-adjoint operator $G_{Rx}$ is Hermitian i.e., it can be written as
\begin{equation}
\label{eq:hermitian}
    (G_{Rx} \psi)(\mathbf{r}_{Rx}) = \int_{\mathcal{S}_{Rx}} g_{Rx}(\mathbf{r}_{Rx} - \mathbf{r}_{Rx}') \psi(\mathbf{r}_{Rx}') d \mathbf{r}_{Rx}'
\end{equation}
where $g_{Rx}(\mathbf{r}_{Rx} - \mathbf{r}_{Rx}')$ is the kernel of the self-adjoint operator.

By direct inspection of \eqref{eq:hermitian}, we evince that $G_{Rx}$ becomes a convolution when $\mathcal{S}_{Rx}$ is asymptotically large. In such a case, the kernel $g_{Rx}(\mathbf{r}_{Rx} - \mathbf{r}_{Rx}')$ acts as a filter in the wavenumber domain. Conceptually, this is analogous to the asymptotic result stated in (14), which highlights, once again, that the definition of the number of eDoF, and hence the dimensionality of the considered space, is legitimate only asymptotically.

Bearing this in mind, the analytical framework introduced in \cite{Landau1975} is briefly summarized in what follows. Let $H_G(\mathbf{k}_{Rx})$ denote the Fourier transform of $g_{Rx}(\mathbf{r}_{Rx})$. Furthermore, let $\mathcal{Q}_{\gamma}$ be the wavenumber support  of $H_G(\mathbf{k}_{Rx})$ corresponding to an accuracy level $\gamma$, which is defined as follows:
\begin{equation}
    \mathcal{Q}_{\gamma} = \{\mathbf{k}_{Rx} \, : \, H_G(\mathbf{k}_{Rx}) \geq \gamma\}
\end{equation}

\begin{figure}[!t]
    \centering 
    \vspace{0.1cm}
    \includegraphics[width=0.75\columnwidth]{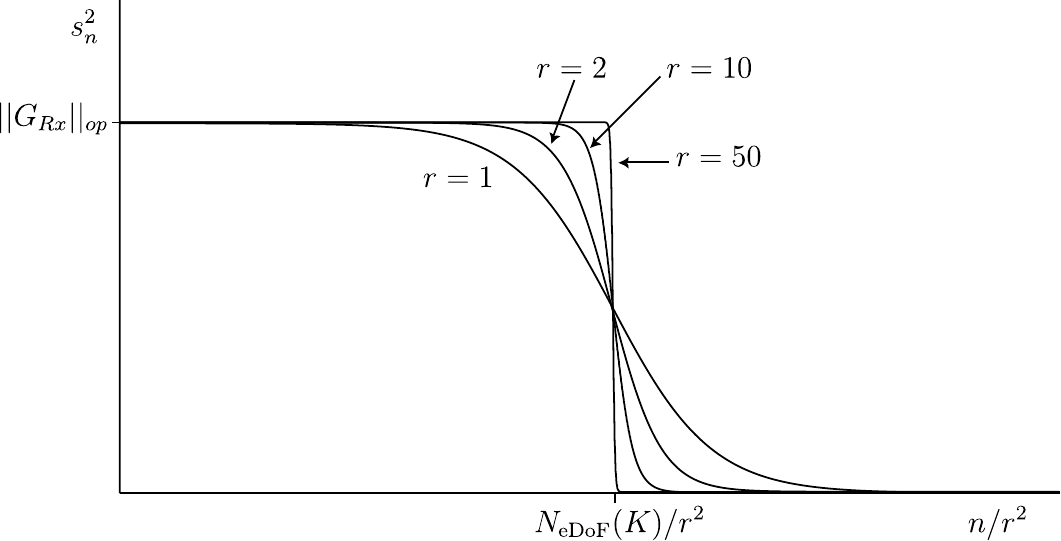}
    \vspace{-0.1cm}
    \caption{Asymptotic behavior of the eigenvalues when $H_G(\mathbf{k}_{Rx}) = ||G_{Rx}||_{op} \mathds{1}_{\mathcal{Q}}(\mathbf{k}_{Rx})$}
    \label{fig:asymtoticRegime} \vspace{-0.5cm}
\end{figure}

To compute the number of \ac{eDoF} in the asymptotic regime, the space support is first expressed as $\mathcal{S}_{Rx}' = r \mathcal{S}_{Rx}$ with $\mathcal{S}_{Rx}$ being a fixed domain. In practice, $\mathcal{S}_{Rx}$ is the actual size of the receiving HoloS, while the parametrization $\mathcal{S}_{Rx}' = r \mathcal{S}_{Rx}$ characterizes the asymptotic regime as $r \to \infty$. Then, Landau's eigenvalue theorem states the following \cite[Th. 2]{Landau1975}:
\begin{equation}
\label{eq:edof_landau}
    \lim_{r \rightarrow \infty} r^{-2} N_{\mathrm{eDoF}, \gamma}(K) = \frac{m(\mathcal{S}_{Rx} )m(\mathcal{Q}_{\gamma})}{(2\pi)^2}
\end{equation}

In contrast to the cut-set integral method, Landau's eigenvalue theorem allows us to compute the number of \ac{eDoF} for any accuracy level $\gamma$. More specifically, the number of eDoF in \eqref{eq:edof_landau} tends to (14) if the wavenumber support is computed for an arbitrarily small value of $\gamma$.

An important case study is when $g_{Rx}(\mathbf{r}_{Rx})$ is an ideal pass-band filter, which can be formulated as follows:
\begin{equation}
\label{eq:idealFilter}
    H_G(\mathbf{k}_{Rx}) = ||G_{Rx}||_{op} \mathds{1}_{\mathcal{Q}}(\mathbf{k}_{Rx})
\end{equation}

Then, for any values of $\gamma$ in $0 \leq \gamma \leq ||G_{Rx}||_{op}$, the number of \ac{eDoF} is $N_{\mathrm{eDoF}, \gamma}(K) = N_{\mathrm{eDoF}}(K)$, where \cite[Th. 1]{Landau1975}
\begin{equation}
\label{eq:LandauTh1}
    \lim_{r \rightarrow \infty} r^{-2} N_{\mathrm{eDoF}}(K) = \frac{m(\mathcal{S}_{Rx} )m(\mathcal{Q})}{(2\pi)^2}
\end{equation}

\begin{table*}[t]
\centering
\caption{Comparison between the analytical methods}
\label{tab:summary}
\begin{tabular}{p{14.5em}||p{22em}|p {19.0em}}
\hline 
\hspace{0.75cm} \textbf{Analytical Method} & \hspace{2.25cm} \textbf{Cut-set Integral}                                                          & \hspace{1.4cm} \textbf{Self-adjoint Operator}                        \\ \hline \hline
Computation of $N_{\mathrm{eDoF}, \gamma} $ & Arbitrarily small $\gamma>0$                                          & For any $\gamma$                            \\ \hline
Regime to analyze the eDoF                   & Asymptotic regime                                                         & Asymptotic regime                            \\ \hline
Applicability to compute the eDoF      & There exist analytical expressions for the local  bandwidth  only for LoS and isotropic channels & The self-adjoint operator needs to be Hermitian \\ \hline
Computation of eigenfunctions                  & Not possible                                                                        & Possible by solving the eigenfunction equation                          \\ \hline
\end{tabular} \vspace{-0.5cm}
\end{table*}

The asymptotic result in (26) implies that the eigenvalues of $G_{Rx}$ polarize asymptotically, i.e., the $N_{\mathrm{eDoF}}(K)$ dominant eigenvalues have a magnitude close to $||G_{Rx}||_{op}$, while the remaining eigenvalues are nearly zero. Figure \ref{fig:asymtoticRegime} shows an illustration of the transition of the eigenvalues when approaching the asymptotic regime. In practice, the asymptotic regime can never be reached, but the framework in (26) is shown to provide a good estimate of the number of eDoF when $N_{\mathrm{eDoF}}(K) \gg 1$ \cite{Pizzo2022}. Conceptually, therefore, this is once again similar to the asymptotic result in (14). If the considered setting does not strictly fulfill the assumptions required by the asymptotic regime, then the transition window of the eigenvalues scales as $o(r^2)$ for bidimensional support sets. The second-order term accounts for the dependence on the geometry of the support set and the accuracy level $\gamma$ \cite{FranceschettiBook}.

\subsection{Application to Holographic MIMO}
Recent research works that applied the self-adjoint operator method to compute the number of \ac{eDoF} of holographic MIMO include \cite{Miller, Pizzo2022, ruizsicilia2023degrees, Do2023a}. In \cite{Pizzo2022}, the authors proved that, for  \ac{LoS} links assuming a paraxial setting (i.e., the size of the \ac{HoloS} is much smaller than the distance between their centers), the wavenumber support can be modeled as in \eqref{eq:idealFilter} and Landau's eigenvalue theorem can hence be applied. In \cite{ruizsicilia2023degrees}, the authors obtained the number of \ac{eDoF} for non-paraxial setups by i) partitioning a large \ac{HoloS} into smaller sub-surfaces in which the paraxial approximation holds true, ii) computing the number of eDOF by using \eqref{eq:LandauTh1} and, finally, iii) integrating the contributions over all the sub-surfaces. The resulting framework boils down, under the same assumptions, to the frameworks in \cite{Dardari_2020, Ding2022}, which are obtained using the cut-set integral approach. The self-adjoint operator was also utilized in \cite{Do2023a} to compute the number of \ac{eDoF} in a communication system aided by a reflecting surface.

It is worth mentioning that Landau's eigenvalue theorems in (24) and (26) characterize the number of \ac{eDoF} of a communication link but do not provide any information on the optimal basis functions to exploit them. If the self-adjoint operator is known, however, it is possible to solve analytically the eigenfunction equation in \eqref{eq:eigen2} in some cases. In \cite{Miller}, the author obtained analytical expressions for the basis functions assuming a paraxial \ac{LoS} setting and a broadside configuration. In \cite{ruizsicilia2023degrees}, the authors generalized the expression of the basis functions for application to some non-broadside settings.

\section{Comparison of Cut-Set Integral and Self-Adjoint Operator}
A comparison between the cut-set integral and self-adjoint operator is presented in Table \ref{tab:summary}. Broadly speaking, the cut-set integral is a general method that can be applied to any channel operator, while the self-adjoint operator method can be directly applied provided that the considered self-adjoint operator is Hermitian. However, analytical expressions for the local bandwidth are available only for some channel operators. In this regard, the self-adjoint operator method may be deemed more flexible, as it can be applied to complex wireless scenarios \cite{Do2023a}. The cut-set integral provides the number of \ac{eDoF} only for an arbitrarily small $\gamma > 0$, while the self-adjoint method can be applied for any values of $\gamma$, offering a characterization of the magnitude of the eigenvalues as well. The cut-set integral method can be applied to compute only the number of \ac{eDoF}, while the self-adjoint operator method provides information on the optimal basis functions as well.

\section{Conclusion}
We presented the two main known methods to compute the number of eDoF for application to holographic MIMO communications: the cut-set integral and the self-adjoint operator. We compared the main benefits and limitations of each approach, and summarized the existing research works.

\section*{Acknowledgment}
This work was supported in part by the European Commission through the H2020 MSCA 5GSmartFact project under grant agreement number 956670.  The work of M. Di Renzo was supported in part by the European Commission through the Horizon Europe project COVER under grant agreement 101086228, the Horizon Europe project UNITE under grant agreement 101129618, and the Horizon Europe project INSTINCT under grant agreement 101139161, as well as by the Agence Nationale de la Recherche (ANR) through the France 2030 project ANR PEPR Networks of the Future under grant agreement NF-FOUND 22-PEFT-0010, and by the CHIST-ERA project PASSIONATE under grant agreements CHIST-ERA-22-WAI-04 and ANR-23-CHR4-0003-01.

\balance

\bibliography{references}
\addcontentsline{toc}{chapter}{Bibliography}
\bibliographystyle{IEEEtran}

% that's all folks
\end{document}